**I. S. Klyuyev\*, D. V. Kasperovych, V. V. Kobychev**

*Institute for Nuclear Research, National Academy of Sciences of Ukraine, Kyiv, Ukraine*

\*Corresponding author: klyuyev218@gmail.com

## SEARCH FOR SPONTANEOUS FISSION OF PLUTONIUM AND AMERICIUM NUCLIDES USING GAMMA SPECTROMETRY

Spontaneous fission in americium and plutonium isotopes was investigated using high-resolution gamma spectrometry. This process was investigated through analysis of a certified $PuO_2$ sample spectrum obtained from the IAEA database of reference spectra. Measurements were carried out with a HPGe detector, and detection efficiency calibration was performed by Monte Carlo simulations using GEANT4. There are no peaks observed that can be ascribed to beta decay of expected spontaneous-fission daughters. Thus, lower experimental limits on the partial half-lives for some spontaneous fission channels in $^{238-242}$Pu and $^{241}$Am were established. In many cases these limits have exceeded current theoretical predictions for cold fission done by S.B. Duarte et al., indicating the need to refine current models for the description of this process.

## 1. Introduction

Spontaneous fission is type of nuclear decay when heavy atomic nucleus divides into two or more lighter fragments, emitting prompt neutrons in the process, without any external influence, with cold fission being one of its possible modes. Cold fission is a rare process of nuclear fission (both spontaneous and induced) for which prompt neutrons are not emitted. Cold fission results in total kinetic energy of fragments that nearly equal the reaction's Q-value, leaving minimal energy for post-scission de-excitation.

In a work by J. Kaufmann et al. [1], a detailed experimental study of thermal neutron-induced fission in $^{232}$U and $^{239}$Pu was performed with the *Cosi fan tutte* spectrometer of the Institut Laue-Langevin in Grenoble. The experimental results indicated that, within the framework of the scission-point model, both compact and deformed scission configurations were observed, in which the resulting fragments carried no intrinsic excitation energy. In [2], the direct evidence of cold fission of $^{242}$Pu was reported using high-resolution gamma-ray coincidence spectroscopy with the GAMMASPHERE array. Zero-neutron emission fragment pairs such as Zr-Xe, Sr-Ba, and Mo-Te were identified.

A unified theoretical model of radioactive decay involving the emission of nucleons and nuclei, developed by S.B. Duarte et al. [3], encompasses a wide range of nuclear transformations – proton emission, alpha and cluster decays, and cold fission. Particularly, this work provides quantitative predictions of partial half-lives for various cold fission channels of heavy elements, including isotopes of plutonium and americium.

## 2. Experimental data analysis

### 2.1. Experimental setup

Experimental spectra from the International Database of Reference Gamma Spectra (IDB) provided by the International Atomic Energy Agency (IAEA) [4] were analyzed in the present work. Spectrum ID 1463 was selected, which includes the metadata with all key experimental parameters and the certified data of the sample used in the measurement [5]. In the experiment, a plutonium oxide ($PuO_2$) pellet with dimensions $\varnothing$ 14.90 × 3.85 mm$^2$ encased in stainless steel and protected with a plastic cap was used. The sample was certified on June 20, 1986, by the Central Bureau for Nuclear Measurements of the Joint Research Centre of the European Commission (EC JRC) in Belgium.

The measurements were carried out on January 19, 2015, using a high-purity germanium (HPGe) detector with a relative efficiency of 50 % [5, 6]. The spectra with and without the sample were measured over a live time of $1.5 \cdot 10^5$ s each, with dead times of 6.61 and 0.02 %, respectively. The sample was placed 22.4 cm from the detector, and cadmium and PMMA attenuators were used with thicknesses of 1 and 6.4 mm, respectively.

Specifications of the complete information about the sample and the detector used for the measurements can be found in Tables 1 and 2.

The isotopic composition of americium and plutonium was decay-corrected from the certification date, June 20, 1986, to the measurement date, January 19, 2015, covering a period of 28.58 years.

*Table 1.* **Properties of the PuO$_2$ sample [7] and measurement conditions [6]**

| Property | Value | |
|---|---|---|
| PuO$_2$ pellet | | |
| Diameter, mm | 14.92 | |
| Thickness, mm | 3.80 | |
| Flatness, mm | < ±0.02 | |
| Mass, g | 6.623 | |
| Density, g·cm$^{-3}$ | 9.969 | |
| Surface density, g·cm$^{-2}$ | 3.80 | |
| Stoichiometry (PuO$_{2-x}$), x | < 0.05 | |
| Impurity level, mg/g | ≤ 0.2 | |
| HPGe detector | | |
| Detector geometry | Coaxial | |
| Detector size | 50 % (rel. eff.) | |
| Detector FWHM, keV | 1.05 | |
| Energy range of the detector, keV | 0.0 - 2927.9 | |
| Source to detector distance, cm | 22.42 | |
| Analyzer name | Lynx (Canberra) | |
| Number of channels | 8 192 | |
| Measurements | | |
| Attenuators | Cadmium, 1 mm | PMMA, 6.4 mm |
| Live time | 150 000 s | |
| Dead time | 6.61 % | |
| Total number of counts | 750 210 853 | |

*Note.* The uncertainty of the mass was unknown, so we assumed it to be negligible.

*Table 2.* **Isotopic composition of the sample (relative to the total content of Pu isotopes)**

| Nuclides | Isotopic composition of the material, at. % | |
|---|---|---|
| | June 20, 1986 [7] | January 19, 2015 (*calculated*) |
| $^{238}$Pu | 0.0117(2) | 0.00933(16) |
| $^{239}$Pu | 93.4392(40) | 93.36244(40) |
| $^{240}$Pu | 6.2886(39) | 6.26964(39) |
| $^{241}$Pu | 0.2215(4) | 0.05558(10) |
| $^{242}$Pu | 0.0390(3) | 0.0390(30) |
| $^{241}$Am | 0.1039(21) | 0.2606(20) |

*Note.* Uncertainties of the last digits are given with 95 % confidence level (C.L.).

The plutonium isotopes were corrected via:

$$N_j(t) = N_j(t_{\text{ref}}) \cdot \exp(-\lambda_j \Delta t), \quad (1)$$

where $\lambda_j = \dfrac{\ln 2}{T_{1/2,j}}$, and $\Delta t \approx 28.58$ years, $N_j$ is the fraction value, $t_{\text{ref}}$ is the reference time point.

For $^{241}$Am, ingrowth from $^{241}$Pu was included leading to:

$$N_{\text{Am}241}(t) = N_{\text{Am}241}(t_{\text{ref}}) e^{-\lambda_{\text{Am}}\Delta t} + $$
$$+ N_{\text{Pu}241}(t_{\text{ref}}) \frac{\lambda_{\text{Pu}}(e^{-\lambda_{\text{Pu}}\Delta t} - e^{-\lambda_{\text{Am}}\Delta t})}{\lambda_{\text{Am}} - \lambda_{\text{Pu}}}. \quad (2)$$

The results (see Table 2) show the expected decrease in $^{241}$Pu and corresponding increase in $^{241}$Am.

### 2.2. Energy and energy resolution calibration

Calibration of the detector's energy scale was performed using well-separated gamma peaks of $^{238,239}$Pu, and $^{241}$Am (+$^{232}$Th series daughters, $^{228}$Ac, $^{208}$Tl, $^{214}$Bi, etc.).

Each peak was fitted with a model function consisting of a linear polynomial to describe the background and a modified left-tailed Gaussian function to approximate the corresponding gamma peak.

$$f(x) = \begin{cases} H \cdot e^{\frac{T(2x-2\mu+T)}{2\sigma^2}}, & \text{for } x < \mu - T, \\ H \cdot e^{-\frac{(x-\mu)^2}{2\sigma^2}}, & \text{for } x \geq \mu - T. \end{cases} \quad (3)$$

Here, $H$ is the peak height, $\sigma$ is the standard deviation of the Gaussian, $T$ is the tailing parameter, and $\mu$ is the centroid of the peak.

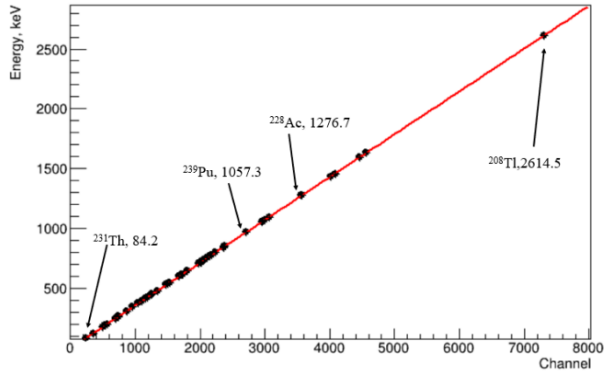

Fig. 1. Energy calibration of spectrum (identification is given for several gamma lines). The linear function obtained by the fit procedure is shown. (See color Figure on the journal website.)

The calibration of the detector energy scale was done by using clear gamma-ray peaks present in the spectrum by the 1st degree polynomial function. The bin width obtained from the calibration curve (Fig. 1) was 0.358 keV.

The energy dependence of peak width and tailing parameters was obtained from the fit procedures, such as:

$$\sigma(\text{keV}) = \sqrt{0.136201(17) + 0.00034272(5) \cdot E_\gamma(\text{keV})}, \quad (4)$$

$$T(\text{keV}) = \sqrt{1.87984(17) + 0.00899663(11) \cdot E_\gamma(\text{keV})}. \quad (5)$$

The spectrum measured with the plutonium sample, together with the detector background, converted into the same energy scale, is shown in Fig. 2.

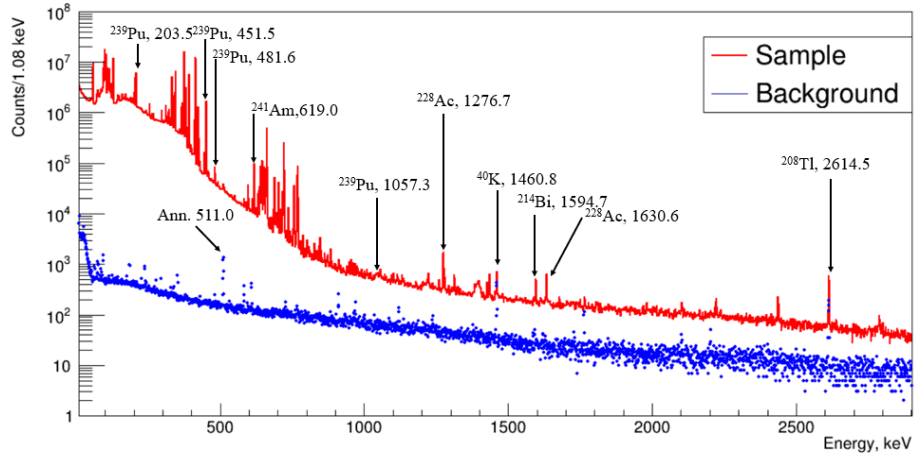

Fig. 2. Energy spectra of the $PuO_2$ sample, acquired with the HPGe detector over 150,000 s (red histogram) and the background spectrum measured over the same live time (blue dots). The gamma-ray peaks are shown with arrows, and the energy is in keV. (See color Figure on the journal website.)

### 2.3. Detection efficiency

Detection efficiency calibration for the full-energy absorption peaks was performed using known gamma peaks of $^{238,239}$Pu, and $^{241}$Am. The detection efficiency $\varepsilon$ for each gamma peak was determined using the following formula:

$$\varepsilon_{exp} = \frac{S}{A \cdot I_\gamma \cdot t}, \quad (6)$$

where $S$ is the area of the full-absorption peak, $A$ is the activity of each parent nuclide in Bq, $t$ is a measurement live time, and $I_\gamma$ is the absolute intensity of each gamma peak. Activity of nuclide $A$ was calculated using data from the IDB database using the corresponding formula:

$$A_i = \frac{\ln 2}{T_{1/2}} \cdot N_i = \frac{\ln 2 \cdot m \cdot \Delta_i \cdot N_A}{T_{1/2} \cdot M}, \quad (7)$$

where $T_{1/2}$ is the half-life value for each nuclide, $N_i$ is the number of nuclei, $m$ is the total mass of the sample, $\Delta_i$ is the isotopic abundance of each nuclide, $M = 271.12$ g/mol is the molar mass of the material, and $N_A$ is the Avogadro's constant. The values of activity, half-lives, molar masses, and the number of nuclei for each nuclide are presented in Table 3.

Energies of gamma lines expected in decays of nuclei which could be created in the spontaneous fission process, however, do not coincide with energies of plutonium and americium gamma peaks. To calculate efficiencies for these energies, a Monte Carlo simulation model was developed using Simourg 1.5.0 [10] code, which employs GEANT4 packages to simulate the experimental geometry [11, 12]. The modeled setup was based on a scheme from the JRC publication repository (Fig. 3), where the spectrum was originally measured [6].

*Table 3.* **Calculated values of activity, half-lives, molar masses [8], and the number of nuclei for each nuclide on January 19, 2015**

| Nuclide | Molar mass, g/mol | Number of nuclei | Half-life, y [9] | Activity, Bq |
|---|---|---|---|---|
| $^{238}$Pu | 238.0495582(12) | 1.37(12)·10$^{18}$ | 87.7(1) | 3.44(30)·10$^8$ |
| $^{239}$Pu | 239.0521616(12) | 1.37(29)·10$^{22}$ | 2.4(3)·10$^4$ | 1.2513(78)·10$^{10}$ |
| $^{240}$Pu | 240.0538117(12) | 9.223(29)·10$^{20}$ | 6561(7) | 3.087(29)·10$^9$ |
| $^{241}$Pu | 241.0568497(12) | 8.176(73)·10$^{18}$ | 14.3(6) | 1.253(28)·10$^{10}$ |
| $^{242}$Pu | 242.0587410(13) | 5.74(22)·10$^{18}$ | 3.73(2)·10$^5$ | 3.38(22)·10$^5$ |
| $^{241}$Am | 241.0568273(12) | 3.83(38)·10$^{19}$ | 432.6(6) | 1.95(20)·10$^9$ |

*Note.* Half-life values were taken from [9], and the activities were calculated using data from the sample certificate [7].

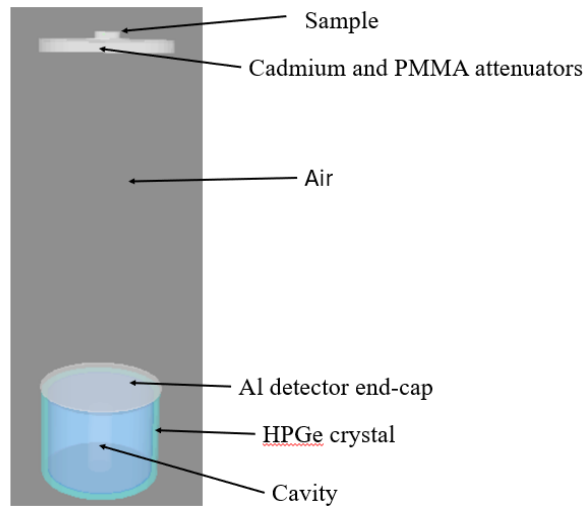

Fig. 3. Measurement geometry built in Simourg 1.5.0. (See color Figure on the journal website.)

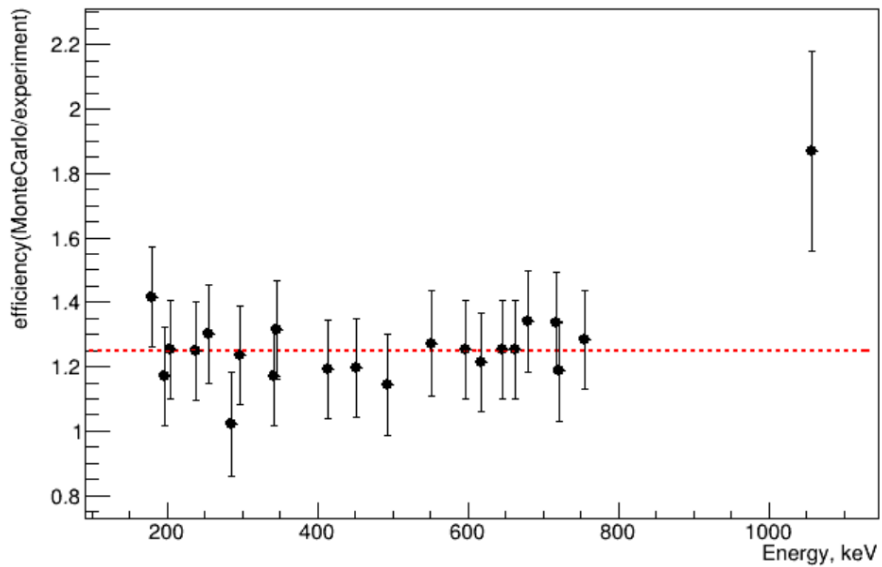

Fig. 4. Dependency of the $\varepsilon_{MC} / \varepsilon_{exp}$ ratio on energy. The approximation function used to fit the data was a constant function, represented by the red dashed line. (See color Figure on the journal website.)

To evaluate how well the simulations describe the experimental values, a $\varepsilon_{MC} / \varepsilon_{exp}$ ratio was used, representing the relationship between experimental and Monte Carlo efficiencies as a function of energy. Then, the obtained dependence was fitted by a constant and shown in Fig. 4.

**2.4. Limits on spontaneous fission half-lives**

In the search for spontaneous fission processes (including the cold fission), we conducted a search for gamma peaks of beta-active secondary nuclides that are considered to be formed in these processes. An example of the search for such peaks is shown in Fig. 5.

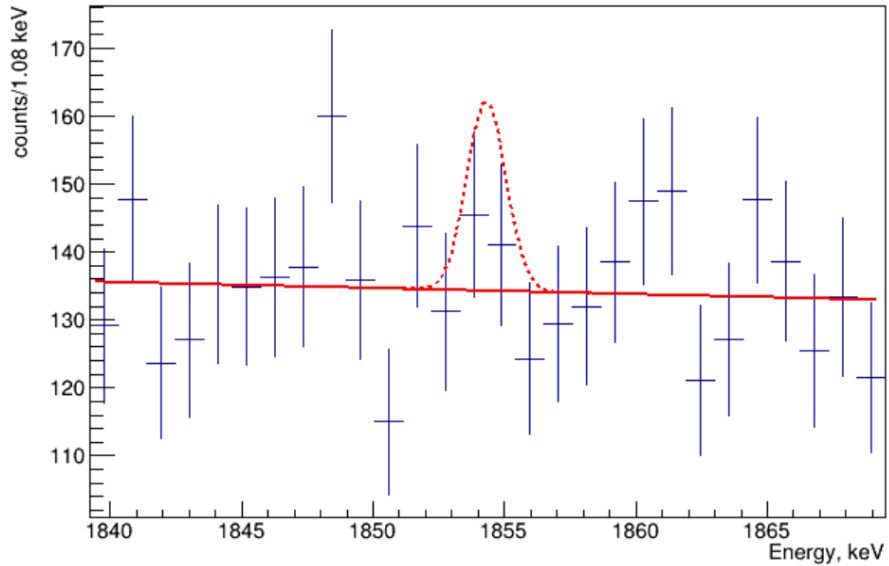

Fig. 5. Example of a search for a peak of $^{131}$Sb (1854.31 keV), which can originate in $^{239}$Pu→$^{108}$Ru + $^{131}$Sn as a daughter of $^{131}$Sn. The background fit model is shown as a solid line, and the peak area lim $S$ = 57 counts, excluded with 90 % C.L. by the Feldman - Cousins procedure [13], is shown as a dotted line. (See color Figure on the journal website.)

No peaks corresponding to spontaneous fission products were observed. Based on the obtained results, lower experimental limits were established for the partial half-lives of several possible spontaneous fission channels of the nuclides $^{238-242}$Pu and $^{241}$Am. To determine the partial half-life limits, the following formula was used:

$$\lim T_{1/2} = \frac{\ln 2 \cdot N \cdot t \cdot \varepsilon \cdot I_\gamma}{\lim S}, \quad (8)$$

where $N$ is the number of parent nuclei in the sample (see Table 3), $t$ is the measurement time (150 000 s) $I_\gamma$ is the absolute intensity of the considered gamma line, ε is the detection efficiency (obtained from the simulations and divided by the correction factor 1.25), and lim $S$ is the upper limit of the gamma peak area, estimated according to the procedure described in [13].

Table 4. **Experimental limits on partial half-lives for spontaneous fission compared to theoretical values for cold fission [3]**

| Nuclide | Spontaneous-fission channel | Reference nuclide and gamma peak, keV | $I_{abs}$, % | ε, % | lim $S$ | Partial $T_{1/2}$, y | |
|---|---|---|---|---|---|---|---|
| | | | | | | Theoretical predictions [3] | This work |
| $^{238}$Pu | $^{104}$Mo + $^{134}$Te | $^{134}$I(1455.2) | 2.3 | 0.042 | 136.5 | 6.8·10$^{11}$ | >3.2·10$^{8}$ |
| $^{239}$Pu | $^{110}$Ru + $^{129}$Sn | $^{129}$Sb(1738.2) | 7.5 | 0.022 | 44.1 | 5.7·10$^{12}$ | **>1.7·10$^{13}$** |
| | $^{108}$Ru + $^{131}$Sn | $^{131}$Te(1146.9) | 4.9 | 0.064 | 84.7 | 2.8·10$^{12}$ | **>1.7·10$^{13}$** |
| | $^{107}$Ru + $^{132}$Sn | $^{132}$I(1372.1) | 2.5 | 0.048 | 15.5 | 1.1·10$^{13}$ | **>3.5·10$^{13}$** |
| | $^{109}$Ru + $^{130}$Sn | $^{130}$Sb(1368.7) | 1.1 | 0.048 | 30.5 | 2.4·10$^{12}$ | **>7.8·10$^{12}$** |
| | $^{111}$Ru + $^{128}$Sn | $^{128}$Sb(1158.2) | 1.5 | 0.063 | 64.8 | 3.1·10$^{13}$ | >6.6·10$^{12}$ |
| | $^{104}$Mo + $^{135}$Te | $^{135}$I(1502.8) | 1.1 | 0.038 | 52.4 | 1.2·10$^{14}$ | >3.5·10$^{12}$ |
| | $^{105}$Mo + $^{134}$Te | $^{134}$I(1455.2) | 2.3 | 0.042 | 136.5 | 1.9·10$^{12}$ | **>3.1·10$^{12}$** |
| $^{240}$Pu | $^{106}$Mo + $^{134}$Te | $^{134}$I(1455.2) | 2.3 | 0.042 | 136.5 | 3.5·10$^{11}$ | >2.2·10$^{11}$ |
| | $^{108}$Ru + $^{132}$Sn | $^{132}$I(1372.1) | 2.5 | 0.048 | 15.5 | 2.2·10$^{11}$ | **>2.3·10$^{12}$** |
| $^{241}$Pu | $^{109}$Ru + $^{132}$Sn | $^{132}$I(1372.1) | 2.5 | 0.048 | 15.5 | 1.3·10$^{11}$ | >2.1·10$^{10}$ |
| $^{242}$Pu | $^{110}$Ru + $^{132}$Sn | $^{132}$I(1372.1) | 2.5 | 0.048 | 15.5 | 4.6·10$^{9}$ | **>1.4·10$^{10}$** |
| $^{241}$Am | $^{108}$Ru + $^{133}$Sb | $^{133}$Te(1722.0) | 0.1 | 0.023 | 30.4 | 3.4·10$^{10}$ | >7.7·10$^{8}$ |
| | $^{109}$Ru + $^{132}$Sb | $^{132}$I(1372.1) | 2.5 | 0.048 | 15.5 | 7.4·10$^{12}$ | >9.7·10$^{10}$ |
| | $^{110}$Ru + $^{131}$Sb | $^{131}$Te(1146.9) | 4.9 | 0.064 | 84.7 | 1.1·10$^{11}$ | >4.7·10$^{10}$ |

*Note*. The limits were determined at 90 % C.L. The values, that exceed the theoretical predictions, are in bold. Here $I_{abs}$ is the absolute intensity of the considered gamma line, ε is the detection efficiency, and lim $S$ is the upper limit of the gamma peak area.

A complete list of the calculated half-life limits is provided in Table 4. It should be noted that the specific daughter nuclides could appear not only in cold fission but also in some other processes (hot fission to heavier nuclides with emission of one or more neutrons, or caused by cosmic ray muons, etc.).

## 3. Conclusions

The gamma spectrum retrieved from the open IAEA database was analyzed in detail. It was measured using a HPGe detector with a live time of 150 000 s. The sample was a well-characterized plutonium oxide ($PuO_2$) certified for isotopic composition of plutonium nuclides and $^{241}$Am. Energy and resolution calibration were carried out using well-known gamma peaks appearing in the spectra, and the detection efficiency was calibrated by Monte Carlo simulations based on the Simourg 1.5.0 program developed using GEANT4, applying a correction factor derived from comparison of simulated and experimental detection efficiencies for the known peaks. No gamma lines were found that could be ascribed to $^{238-242}$Pu or $^{241}$Am spontaneous-fission products. Therefore, lower limits on the partial half-lives were set for these processes. In many cases, the limits considerably exceed theoretical predictions for cold fission channels [3]. Therefore, the theoretical model [3] does not accurately describe the cold fission process and requires revision. Further improvements to the model of detection efficiency and systematics consideration will allow refinement of the obtained limits.

**І. С. Клюєв\*, Д. В. Касперович, В. В. Кобичев**

*Інститут ядерних досліджень НАН України, Київ, Україна*

\*Відповідальний автор: klyuyev218@gmail.com

## ПОШУК СПОНТАННОГО ПОДІЛУ НУКЛІДІВ ПЛУТОНІЮ ТА АМЕРИЦІЮ З ВИКОРИСТАННЯМ ГАММА-СПЕКТРОМЕТРІЇ

Спонтанний поділ ізотопів америцію та плутонію досліджувався за допомогою гамма-спектрометрії високої роздільної здатності. Цей процес вивчався шляхом аналізу спектра сертифікованого зразка $PuO_2$, отриманого з бази даних еталонних спектрів МАГАТЕ. Вимірювання проводилися за допомогою HPGe-детектора, а калібрування ефективності реєстрації здійснювалося методом Монте-Карло з використанням пакета GEANT4. У спектрі не виявлено піків, які можна було б віднести до бета-розпаду очікуваних продуктів спонтанного поділу. Таким чином, були встановлені нижні експериментальні межі для парціальних періодів напіврозпаду деяких каналів спонтанного поділу у $^{238-242}$Pu та $^{241}$Am. У багатьох випадках ці межі перевищили наявні теоретичні оцінки холодного поділу, зроблені С.Б. Дуарте та ін., що вказує на необхідність удосконалення сучасних моделей для опису цього процесу.

*Ключові слова*: плутоній, америцій, спонтанний поділ, HPGe-спектрометрія, GEANT4, період напіврозпаду.